\documentclass[a4paper]{spie}  

\usepackage{amsmath,amsfonts}
\usepackage{array}
\usepackage[caption=false,font=normalsize,labelfont=sf,textfont=sf]{subfig}
\usepackage{cite}

\usepackage{tikz}
\usepackage{pifont} 
\usepackage{multirow}
\usepackage{makecell}
\newcommand{\cmark}{\ding{51}}%
\newcommand{\xmark}{\ding{55}}%

\title{A Machine Learning Approach Towards SKILL Code Autocompletion}

\author[a,b]{Enrique Dehaerne}
\author[b]{Bappaditya Dey}
\author[a]{Wannes Meert}
\affil[a]{Dept. of Computer Science, KU Leuven, 3001 Leuven, Belgium}
\affil[b]{Interuniversity Microelectronics Centre (imec), 3001 Leuven, Belgium}

\authorinfo{Send correspondence to Enrique Dehaerne (enrique.dehaerne@kuleuven.be)}

\begin{document}
\maketitle

\begin{abstract}
As Moore's Law continues to increase the complexity of electronic systems, Electronic Design Automation (EDA) must advance to meet global demand. An important example of an EDA technology is SKILL, a scripting language used to customize and extend EDA software. Recently, code generation models using the transformer architecture have achieved impressive results in academic settings and have even been used in commercial developer tools to improve developer productivity. To the best of our knowledge, this study is the first to apply transformers to SKILL code autocompletion towards improving the productivity of hardware design engineers. In this study, a novel, data-efficient methodology for generating SKILL code is proposed and experimentally validated. More specifically, we propose a novel methodology for (i) creating a high-quality SKILL dataset with both unlabeled and labeled data, (ii) a training strategy where T5 models pre-trained on general programming language code are fine-tuned on our custom SKILL dataset using self-supervised and supervised learning, and (iii) evaluating synthesized SKILL code. We show that models trained using the proposed methodology outperform baselines in terms of human-judgment score and BLEU score. A major challenge faced was the extremely small amount of available SKILL code data that can be used to train a transformer model to generate SKILL code. Despite our validated improvements, the extremely small dataset available to us was still not enough to train a model that can reliably autocomplete SKILL code. We discuss this and other limitations as well as future work that could address these limitations.
\end{abstract}


\section{Introduction}\label{actual_intro}
Electronic systems, everything from microwaves to supercomputers, must be designed by hardware engineers. Electronic design is the process of designing an electrical system that meets certain specifications. Electronics can be combinations of connected transistors as well as more complex logical functions, such as processors, controllers, and memory \cite{bebop_boolean}. Electronic design automation (EDA) refers to technologies that automate the design, analysis, and verification of electronic systems. In practice, the purpose of the EDA industry is to provide software tools for hardware engineers to increase their productivity.

Increasing demand and complexity continue to drive innovation in EDA. In 1965, Moore observed that the number of transistors integrated on a chip approximately doubled every two years \cite{moores_law} and predicted that this trend would continue. The ever-increasing complexity of electronic systems requires constant innovation in the EDA industry. One of the first books to advocate for the use of programming languages (PLs) for EDA was \textit{Introduction to VLSI Systems} \cite{intro_to_vlsi}, published in 1980. Since then, many Hardware Description Languages (HDLs) and other PLs for EDA have been developed and continue to be used today.

SKILL is a PL developed by Cadence \cite{cadence_company} that allows hardware design engineers to customize and interact with design software in a programmatic manner \cite{skill_paper}. Notably, SKILL can be used for physical layout design \cite{cadence_virtuoso} as well as printed circuit board design \cite{cadence_allegro} suites. It aims to provide an easy-to-use abstraction layer that is compatible with different underlying design technologies. The creation of SKILL was motivated by the EDA software needs such as computational efficiency and flexibility to underlying design algorithms \cite{skill_paper}.

\begin{figure}
    \centering
    \includegraphics[width=\linewidth]{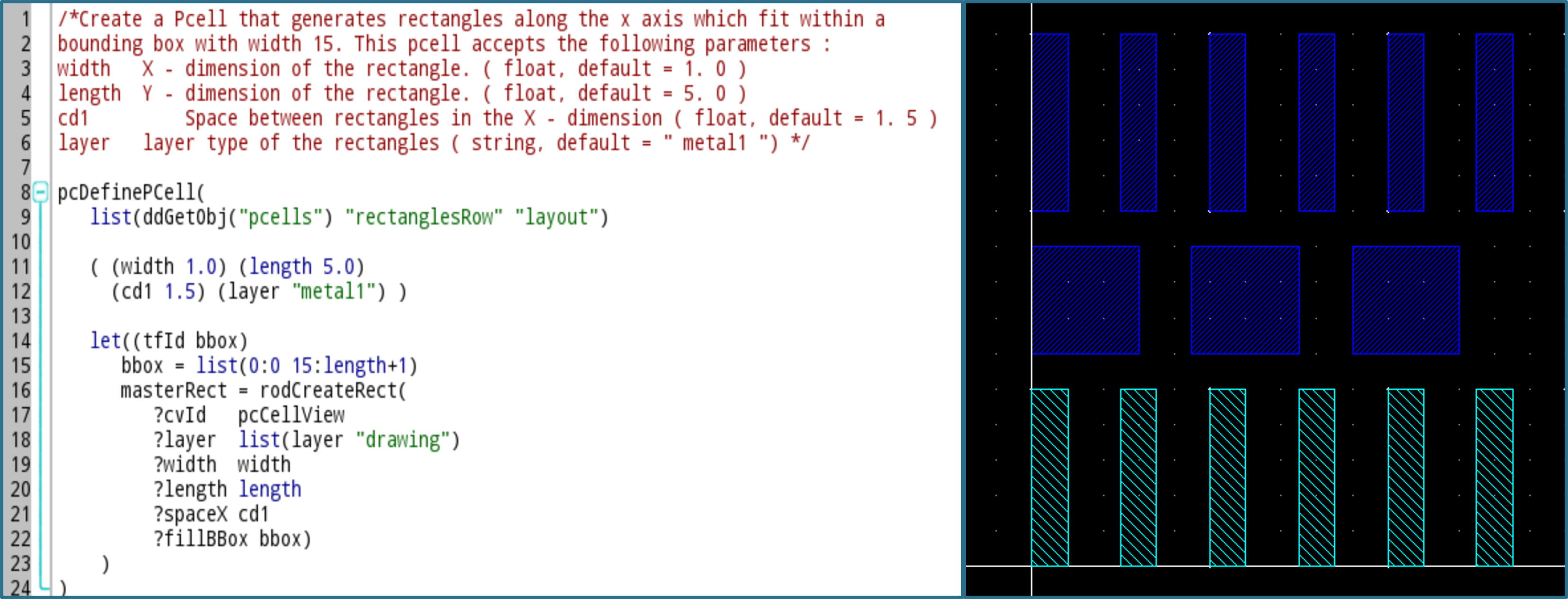}
    \caption{On the left, an example of a parameterized cell (PCell) in the SKILL integrated design environment. The code in red is comments which describe the code underneath it. On the right, instantiations of the PCell with different values of parameters are shown as they appear in the Virtuoso layout editor \cite{cadence_virtuoso}. The top instantiation uses all default parameter values, the middle instantiation has the height and width values set to 3, and the bottom instantiation has the layer type set to ``metal2''.}
    \label{fig:skill_run_example}
\end{figure}

An example SKILL program that describes a parameterized cell (PCell) that can be used to instantiate a (part of a) physical layout is shown on the left side of Figure \ref{fig:skill_run_example}. The code in red is a code comment that describes the PCell and its parameters. When instantiated with the default parameters, the program describes the layout shown in the top right of Figure \ref{fig:skill_run_example}. Below are two instantiations where the values of the parameters differ from the default values.

Machine learning (ML) applied to various steps of the electronic design process has become an important topic of research \cite{ml_for_eda_survey}. Automatic code generation using ML promises to increase the productivity of software developers. Despite a growing body of work in the field of automatic code generation, no work has been published applying this knowledge to SKILL. The transformer \cite{attention_is_all} neural network architecture has become popular in the field of natural language processing for its efficiency in learning sequences of strings. PLs are often represented as strings called source code and can therefore be processed similarly to natural language (NL), that is, by tokenizing strings into sub-strings and modeling sequences of these tokens. In this way, transformers have been trained on huge code datasets to be able to generate competition-level code at an average human level \cite{alphacode} and for commercial code autocompletion tools \cite{copilot}.

In this study, a novel methodology for generating SKILL code is proposed. The goal is to use ML methods to generate SKILL code given code contexts with or without NL descriptions (this task is referred to as autocompletion in the rest of this paper) to help make SKILL developers, and hardware design engineers as a result, more productive. The main challenge to overcome towards completing this goal is the extremely limited amount of available SKILL data. To address this problem, our proposed methodology includes:
\begin{enumerate}
    \item Maximizing the usage of available SKILL code by mining unlabeled and labeled data. In our case, unlabeled data are SKILL code file texts while labeled data are pairs of SKILL code definitions or descriptions and their corresponding bodies. We improve the quality of our dataset through filtering strategies and deduplication.
    \item Transfer learning from models pre-trained on large datasets of general PL data, such as Python and Java code. We choose T5-based \cite{t5} pre-trained transformers because they are designed for transfer learning and can be fine-tuned using both unlabeled and labeled data.
\end{enumerate}
Additionally, we develop a novel, practical methodology for evaluating synthesized SKILL code. This study provides discussion and experimental results towards reliable SKILL code autocompletion. Experimental results show that the proposed models outperform similar baseline models. However, limitations of this study, such as the limited available dataset and model size used, contributed to the low overall performance of the proposed models. Therefore the main contributions of this study are extensive discussions of our (i) proposed methodology for curating a custom SKILL code dataset and fine-tuning a pre-trained T5-based model to autocomplete SKILL code, (ii) its experimental results, and (iii) its limitations. Additionally, we suggest future work to improve our results.

The rest of the paper is organized as follows. Next, Section \ref{new_intro} provides an overview of lessons learned from related work in the field of automatic code generation. Section \ref{methodology} explains the methodology used for (a) creating a custom SKILL dataset (b) training models to generate SKILL code, and (c) evaluating synthesized SKILL code. Section \ref{exp_setup} describes the experimental setup and pre-processing steps used for the experiments. Next, Section \ref{results_discussion} presents and discusses the experimental results of the proposed methodology compared to baselines. Section \ref{limitations} discusses limitations of the our study and suggests future work to address these limitations.

\section{Related Work}\label{new_intro}
SKILL-related works propose manually written SKILL programs that automate layout design. Tayenjam et al. \cite{skill_pcell_inductor_2017} propose a SKILL PCell to automatically generate inductor layouts given certain input parameters. Abhishek et al. \cite{skill_inductor_optimization_2018} propose an algorithm written in SKILL to optimize inductor layouts toward minimizing loss factors. Searches were conducted in an attempt to find code generation-related work specific to the SKILL PL \cite{skill_paper} but none were found. Instead, in the rest of this section, we discuss works related to the automatic generation of HDL code since they are the most similar type of language to the SKILL PL.

Most of the related work on automatic HDL code generation is not ML-based but rather heuristics- or rule-based. The most studied HDLs are Verilog and VHDL. Several research works propose methods that automatically generate Verilog code \cite{generating_fpga_cnn_verilog, Automated_ML_Hardware_Synthesis_verilog} or VHDL code \cite{uml_to_vhdl_2010} from high-level descriptions. Other research works automatically translate code written in other PLs to Verilog \cite{verilog_from_cpp,pythonverilog} or VHDL \cite{c_to_vhdl_2007}. These methods that are manually programmed have the advantage that the generated HDL code is guaranteed to be compilable. The disadvantage of these approaches is that they can only generate HDL code for specific functionalities from specific inputs (a list of parameters, code written with specific APIs, etc).

ML-based code generation methods can generate diverse code snippets from a wide variety of input data, for example, previous-written code \cite{watson2020assert,korbak2021controlling} or NL documentation \cite{chen2021evaluating,hong2021latent}, relatively efficiently. However, ML-based code generation methods generally cannot guarantee the compilability or functional correctness of their generated code. Transformer-based language models \cite{attention_is_all}, that process source code as a sequence of tokens, have become especially popular and effective in code generation tasks \cite{dehaerne2022code_generation_survey}. The primary limitation of token-based language modeling is the need for a large amount of source code data and computations to train the models on this data. Most code generation models are trained on huge datasets of code written in general PLs such as Python \cite{korbak2021controlling} or Java \cite{hassan2020} which are abundantly available on open-source repository databases \cite{dehaerne2022code_generation_survey}. State-of-the-art models are usually trained on a dataset of many different PLs, including the target PL, to maximize the amount of code data the model can learn from \cite{codetrans, codet5, codegen}.

To the best of our knowledge, only two previous works have used ML to generate HDL code. Pearce et al. \cite{dave_verilog} started from the GPT-2 transformer model \cite{gpt2} that was pre-trained on natural language (NL) data and fine-tuned on a synthetic dataset of pairs of English descriptions with Verilog code snippets. Thakur et al. \cite{thakur2022benchmarking} benchmarked recent transformer models pre-trained on general PL code for solving Verilog programming challenges. They fine-tuned the pre-trained models on unlabeled Verilog data from open-source repositories. They manually created a collection of Verilog problems and corresponding test benches to functionally evaluate Verilog code generated by the models.

\section{Methodology}\label{methodology}
This section introduces the proposed methodology for autocompleting SKILL code from NL documentation and/or previously written SKILL code. As discussed in the previous section, the related literature suggests that transformer ML models are best suited for this problem statement. The main challenge that we had to overcome for this research was the extremely limited amount of available SKILL code data. To overcome this challenge, we propose fine-tuning models that are pre-trained on general PL code data through both self-supervised and supervised learning on a custom, curated SKILL code dataset. We evaluate our proposed models and similar baseline models using a SKILL-specific BLEU score as well as the SKILL lint tool. Subsections \ref{dataset_method}, \ref{t5_method}, and \ref{sect:evaluation} discuss the custom SKILL dataset, training of the models, and evaluation of the models, respectively.

\subsection{Custom SKILL Dataset}\label{dataset_method}
 A custom SKILL dataset was created because there are no publicly available SKILL code datasets. Although a few proprietary SKILL repositories were available for this study, the volume of code in these datasets was small. Therefore, open-source repositories were mined for additional SKILL programs to be used to train the transformer models. As will be discussed in Section \ref{results_dataset}, the amount of open-source SKILL data was also small and this motivated the use of models pre-trained on large volumes of general PL data.
 
\begin{figure}[ht]
    \centering
    \includegraphics[width=0.5\textwidth]{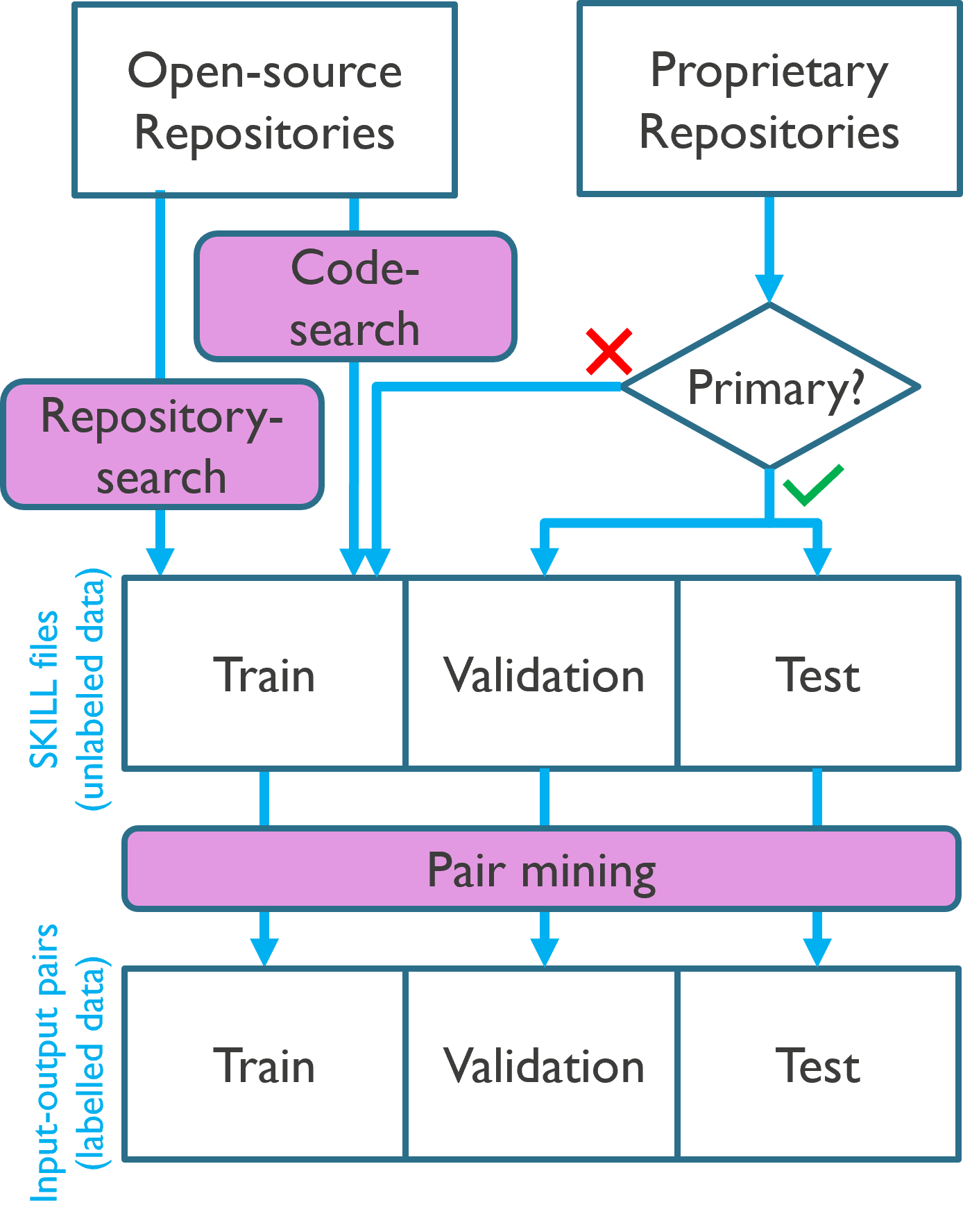}
    \caption{Flowchart showing the sources used and steps taken to create self-supervised and supervised SKILL data. Note that a small amount of data from primary proprietary sources was added to the training split (see Section \ref{filtering}) which is not depicted in this flowchart.}
    \label{fig:dataset_flowchart}
\end{figure}

 First, a collection of SKILL source code files was gathered from the proprietary and open-source repositories. These files were then automatically mined to obtain input-output pairs for supervised training and evaluation. An overview of the methodology used to create the custom SKILL dataset is shown in Figure \ref{fig:dataset_flowchart}. Additionally, three data filtering techniques and one deduplication technique to improve the quality of the training dataset are introduced at the end of this section.

    \subsubsection{Proprietary SKILL Data}\label{sect:prop_data}
    A variety of proprietary SKILL repositories and resources available to us were gathered for the custom SKILL dataset. This proprietary data is considered to be reliable and high-quality since we can verify that they were developed by SKILL experts at imec. We further distinguish proprietary data into two categories, \textit{primary} and \textit{secondary} data. Primary data is data from certain proprietary sources that have desirable qualities for the evaluation of SKILL code autocompletion. We exclusively use primary proprietary data for the validation and test splits of our dataset. Three factors were taken into consideration when classifying a source as primary or secondary.
   
    The first factor was whether the SKILL code was well-documented. The more documentation is available, the more supervised comment-code pairs can be mined out of the data. Code shared between many teams was mostly well documented while those that directly implement solutions (PCells, GUI scripts, etc) were mostly undocumented. An important subset of the code not belonging to one particular team was code with example programs that are used mostly for training SKILL developers.
    
    The second factor was the use of the standard SKILL libraries. This is correlated with the documentation of the files since code shared between teams usually does not use custom libraries while direct implementations do. Adding code contexts from imported libraries to improve the generated code requires inter-file knowledge and linking techniques or advanced code-searching techniques. This is considered out of the scope of this study. That is why, for evaluation purposes, SKILL files that use mostly the standard SKILL libraries were considered to be primary data.
    
    The last factor was the technique used to gather data from a source. Most of the proprietary data were collected simply through zipped directories which required no rule-based extraction or modifications. Data that required rule-based extraction included web-based documentation with program examples. The Selenium WebDriver \cite{selenium_webdriver} package for Python \cite{python_software} was used to mine these types of files. A script with custom rules for the documentation structure was written to extract as many comment-code pairs, saved as SKILL files, from the documentation. A limited number of other SKILL files were modified to make them more homogeneous with the rest of the dataset. These mining rules and code transformations were based on custom heuristics with no guarantee that it works as intended for every sample. Therefore, this data was considered to be secondary data. 
    
    \subsubsection{Open-source SKILL Data}\label{sect:github}
     In addition to the high-quality secondary proprietary data, SKILL code training data was obtained from open-source repositories.
     GitHub \cite{github_website} is popular for collecting large volumes of code data \cite{codenn,public_git_archive,codesearchnet,chen2021evaluating}. Unlike proprietary data, open-source code is not reliably high-quality. Open-source data is therefore only included in the training split of the SKILL dataset, not in the evaluation splits. In Section \ref{filtering}, we propose filtering strategies to remove poor-quality code from our training data. Using the GitHub application programming interface (API), two types of searches were conducted to retrieve publicly available SKILL code. The first search was a repository search. This search retrieved repositories that contain the query phrase: ``cadence skill''. Not including ``cadence'' in the query phrase retrieved many repositories that do not relate to the SKILL language. The repositories were forked and all files with SKILL extensions (\textit{.il} and \textit{.ils}) were added to the custom SKILL dataset. 
    
    Not all SKILL repositories contained the aforementioned query phrase which is why code file searches were also conducted. Frequently occurring tokens from the previously collected proprietary and repository search SKILL files were used as query phrases. Since there are relatively strict limitations on the rate of requests free GitHub users can make via the API, a small subset of all these tokens were searched. More specifically, one-fifth of all tokens that were seen more than 10 times, a total of 1436 tokens, were used as queries for GitHub code searches. Additionally, the file extension filter was used to only retrieve \textit{.il} and \textit{.ils} files. 

    The code file searches retrieved files of other PLs that use the same extensions (e.g., the \textit{Intermediate Language} \cite{intermediate_language}). A list of blacklisted terms that were found to be popular in non-SKILL files and very rare in SKILL files was used to identify and remove non-SKILL files. The files retrieved were first filtered based on the inclusion of several terms that were frequently found in URLs for non-SKILL files. The remaining files were downloaded and further filtered if they contained any of a list of patterns popular in non-SKILL files that are not-syntactically correct in SKILL. The list of exclusion terms for the URLs and exclusion patterns for code are shown in Table \ref{tab:code_search_filtering}.

\begin{table}[ht]
    \caption{Keywords and patterns used to find and remove URLs and files, respectively, retrieved from code searches.}
    \label{tab:code_search_filtering}
    \centering
    \begin{tabular}{|c||>{\raggedright\arraybackslash}m{280pt}|}
    \hline
         \textbf{Filtering stage} & \textbf{Keywords/Patterns} \\\hline\hline
         URL & `dotnet', `-ms', `microsoft', `.net', `solaris', `unity', `logs', `www' \\ \hline
         File & `.assembly', `.NET', `.class', `.method', `.string', `.float', `.inline' (for all: preceded by at least one whitespace character) \\ \hline 
    \end{tabular}
\end{table}
    
\subsubsection{SKILL Input-Output Pairs}\label{sect:supervised_dataset}
    The SKILL file dataset collected from proprietary and open-source repositories was mined to obtain input-output pairs for supervised training and evaluation. Ideally, the inputs give enough information, in the form of comments and/or preceding code, which would prompt a SKILL developer to write code similar to the output reference. This section explains the heuristics used to mine such input-output pairs from file-level data.
    
\begin{figure}[ht]
    \centering
    \includegraphics[width=\textwidth]{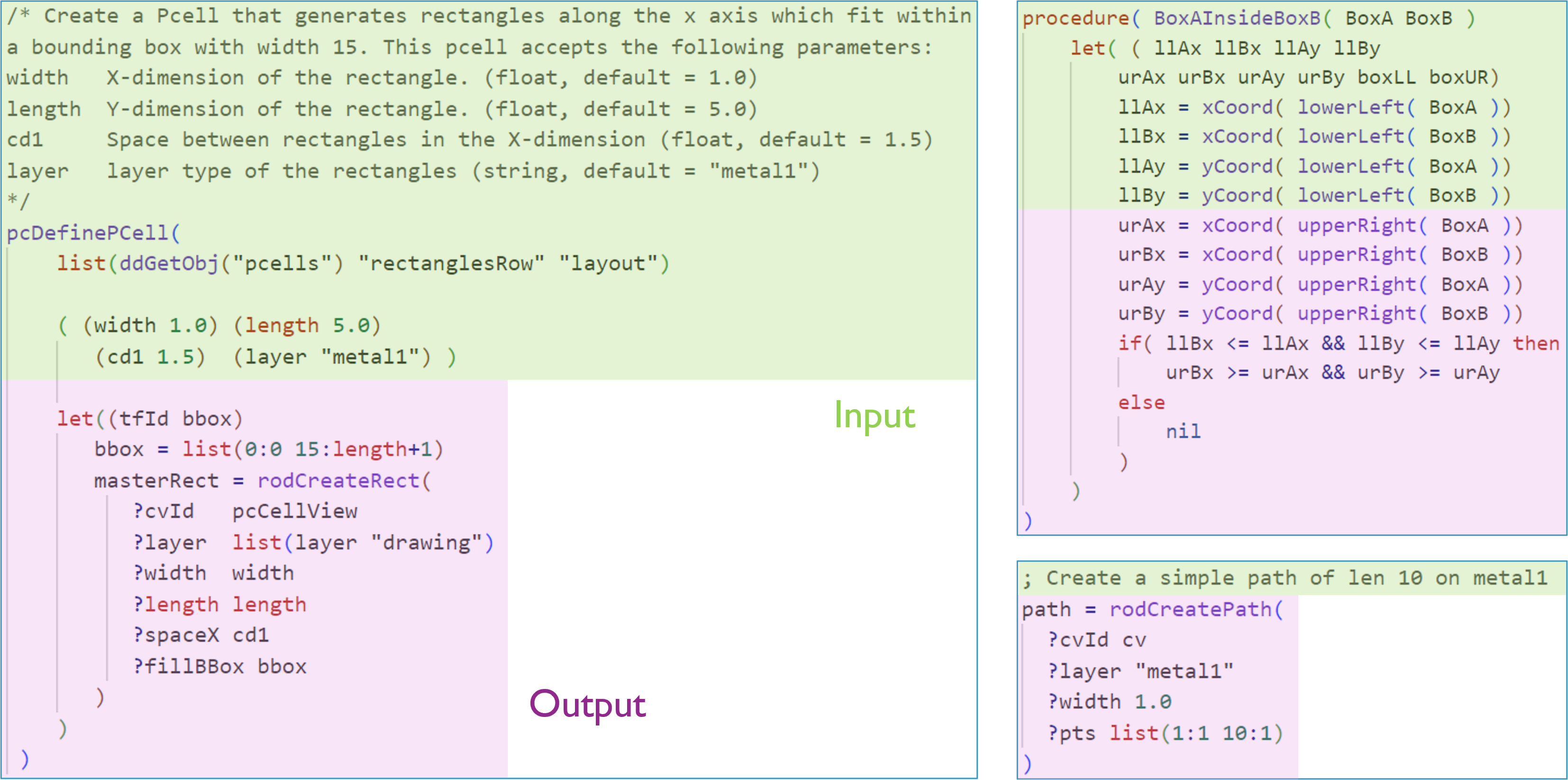}
    \caption{Three example SKILL programs with annotations showing which parts of the program would belong to the input (green) and output (purple) of a pair. The example on the left is equivalent to the program shown in Figure \ref{fig:skill_run_example} and can be split into a comment-function pair (see Section \ref{sect:supervised_dataset}). The top-right and bottom-right pairs are function-completion and comment-code pairs, respectively. Note that these SKILL programs were manually written and were not included in the SKILL dataset.}
    \label{fig:example_input_output}
\end{figure}

    The granularity level for input-output pairs is an important factor. Public datasets for code generation in other PLs often use function-level samples \cite{codesearchnet, deepcom, leclairdataset, parallel_python_dataset}. This is because functions encapsulate code into units of functionality. Additionally, they provide well-defined arguments that are expected to be used in the body, giving the models more information to guide code generation. Function definitions and preceding comments are the input and the function body is the output for \textit{comment-function} pairs. When there are no preceding comments, these definition-body pairs, called \textit{function-completion} pairs, are also mined. To provide additional context from long function bodies, the input in a function-completion pair can include the first parts of the function body with the output being the rest of the body. This was done for function with long bodies. The final pair type mined was \textit{comment-code} pairs. The input of a comment-code pair is a comment and the output is code that directly follows the comment and does not define a function output. This code can be a single code statement or a construct, such as a \textit{foreach} loop, which contains multiple statements. Figure \ref{fig:example_input_output} shows an example for each of the three pair types.
    
\subsubsection{Dataset Filtering, Deduplication, \& Split}\label{filtering}
Different filtering strategies for removing low-quality files and improving learning were experimented with to obtain different data subsets. These strategies were: (i) the file filtering technique used (none, a SKILL lint pass grade, a SKILL lint IQ score greater than or equal to 10, or having more than 1 input-output pair found in the file) and (ii) whether comments were removed from the files. The SKILL lint tool is a static analysis tool used to measure code quality and compilability. It reads files and assigns an IQ score (out of 100) as well as a pass or fail grade to the file. The pass or fail grade is based on significant syntactic errors while IQ takes into account style and efficiency in addition to syntactic correctness. Not training in a self-supervised manner was also experimented with. 

For each data subset that could be obtained by combining the filtering strategies described above, supervised training was either applied or not applied. When applied, whether the pairs were deduplicated was treated as an additional filtering strategy. A pair was considered to be a duplicate of another pair if it appeared within the input and/or output of the other pair. Pairs that are not found within other pairs were called \textit{top-level} pairs. The deduplicated training set only contained top-level pairs. In the rest of this paper, we use the term \textit{training strategy} to denote a possible combination of self-supervised and/or supervised learning with different filtering and deduplication techniques applied to the training dataset.

The validation and testing sets contained a balanced number of comment-function and function-completion pairs (all top-level) as well as comment-code pairs (not all top-level). To achieve a balanced number of each pair type for the test and validation datasets, the following steps were taken:
    \begin{enumerate}
        \item The primary files were randomly partitioned into two sets until a similar number of comment-function pairs were mined from each partition.
        \item The number of function-completion pairs was larger than the number of comment-function pairs ($n$) so $n$ random samples from these pairs were kept and the rest were discarded.
        \item The number of top-level comment-code pairs was smaller than $n$ so the comment-code pairs in each split consisted of all top-level comment-code pairs and adding random samples from non-top-level comment-code pairs until $n$ comment-code pairs were obtained. 
    \end{enumerate}
Note that input-output pairs were used to split files between validation and test splits. This means the files from primary proprietary sources that did not contain any input-output pairs were instead added to the training split.
   
\subsection{Models \& Training}\label{t5_method}

Transformer models \cite{attention_is_all} require large amounts of data to perform well on complex tasks such as code generation. The intuition was that the custom SKILL dataset was too small to train a transformer model from scratch. Therefore, using models with weights pre-trained on general PL code data was hypothesized to result in better final models. The T5 \cite{t5} framework was designed for transfer learning from large unlabeled datasets to task-specific, labeled datasets. CodeTrans \cite{codetrans} and CodeT5 \cite{codet5} are two state-of-the-art models based on T5 \cite{t5} that are trained on a variety of different tasks in a variety of different PLs. Checkpoints of both these models trained on large volumes of code in multiple PLs are publicly available in the HuggingFace library \cite{huggingface}. The CodeTrans model is trained on more data than CodeT5 but CodeT5 uses code-specific learning objectives to improve its learning. To get the most out of our extremely limited SKILL dataset, we extended T5's transfer learning approach by fine-tuning these models on both unlabeled and labeled SKILL data. 

Two baselines were trained and evaluated as well to compare the results to these proposed models. The first baseline is the original T5 model which was trained on a large collection of NL data. This model is called \textit{T5-NL} in the rest of the paper. The second baseline, \textit{T5-SKILL}, is the original T5 model with random initial weights and is therefore only trained on SKILL data. An advantage of training a model from scratch in this way is that the tokenizer can be adapted to the training data. Therefore, a SentencePiece  \cite{sentencepiece} tokenizer was trained on the training set of the custom SKILL dataset which means the T5-SKILL model's vocabulary is SKILL-specific. SentencePiece \cite{sentencepiece} is a sub-word tokenization algorithm that is also used for the CodeTrans \cite{codetrans} and T5-NL \cite{t5} models. CodeT5 uses a byte-pair-encoding \cite{bpe} sub-word tokenizer similar to the one used by GPT3 \cite{gpt3}. The tokenizers for these models are trained on their respective pre-training datasets and are therefore not SKILL-specific.

\begin{figure}
    \centering
    \includegraphics[width=0.7\textwidth]{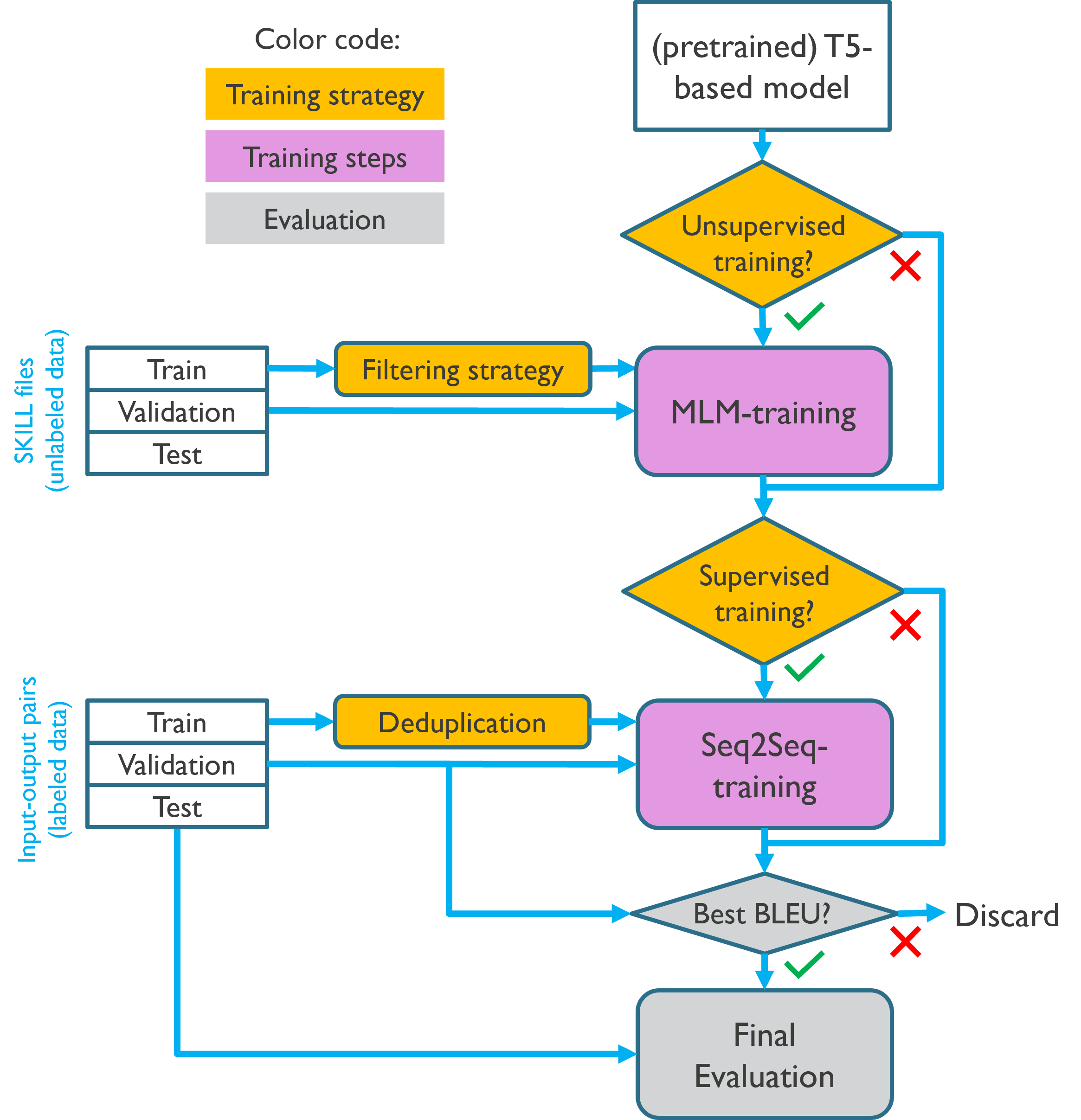}
    \caption{Flowchart showing the high-level training and evaluation steps taken. The ``Best BLEU'' condition is what decides, for each model type, which training strategy resulted in the best-trained model. This model that achieved the best BLEU score was chosen for final evaluation. Note that certain models that did not achieve the best BLEU score for a given model type were also selected for final evaluation (see Section \ref{results_training}).}
    \label{fig:training_flowchart}
\end{figure}

\begin{figure}
    \centering
    \includegraphics[width=0.8\textwidth]{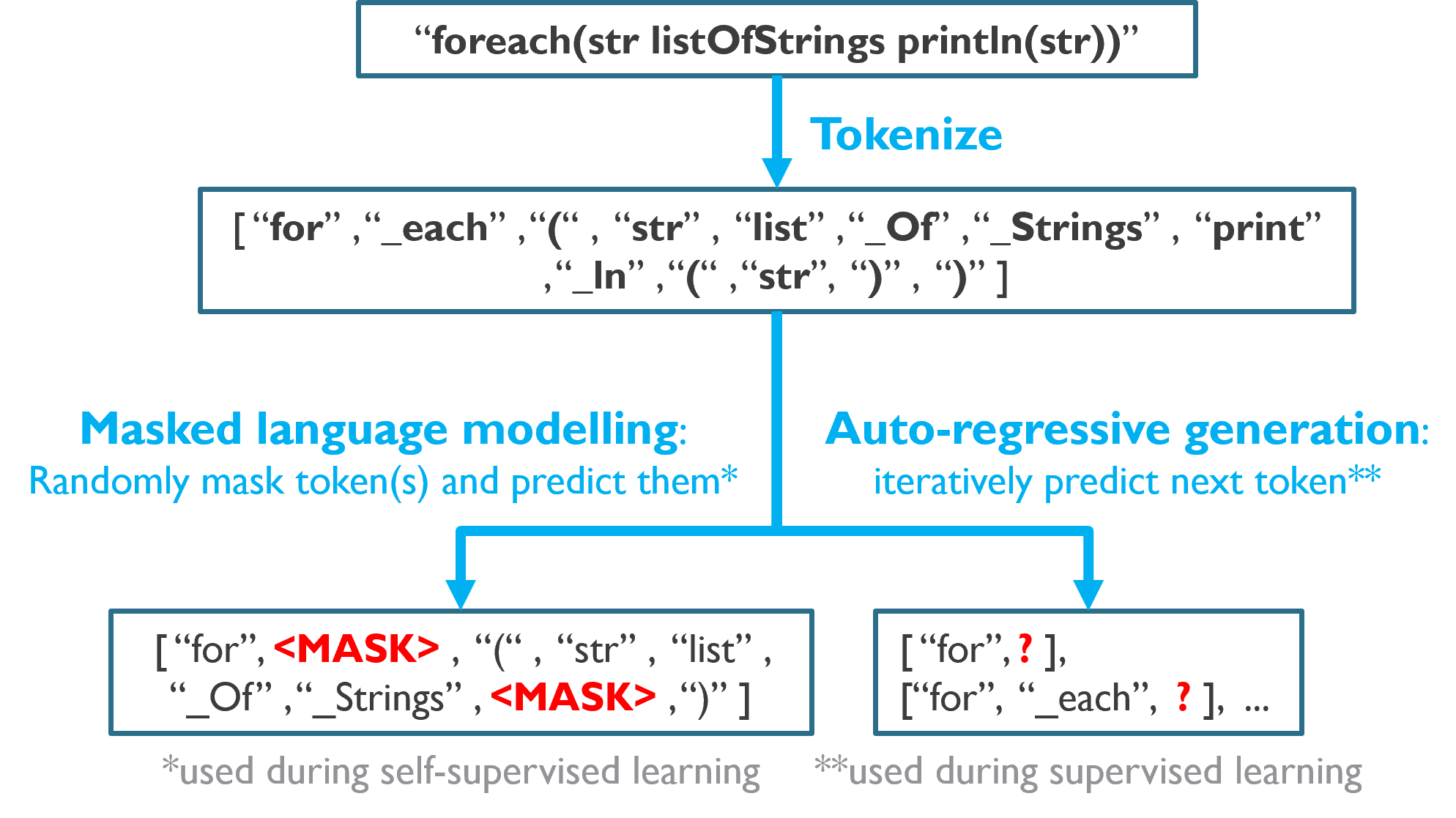}
    \caption{Graphical depiction of MLM and autoregressive modeling for an example SKILL statement. 
    }
    \label{fig:training_objectives}
\end{figure}

The proposed models and both baselines were trained using various training strategies (see Section \ref{filtering}) as shown in Figure \ref{fig:training_flowchart}. For each model type (CodeTrans \cite{codetrans}, CodeT5 \cite{codet5}, T5-NL \cite{t5}, and T5-SKILL \cite{t5}) a model is trained for each of the different possible training strategies. Self-supervised learning is performed using Masked-Language Modeling (MLM) on the SKILL file samples. MLM randomly masks tokens in a token sequence. The task of the ML model being trained is to predict the tokens that have been masked. This allows the model to learn the relationship between tokens before and after the masked tokens. This bi-directional way of modeling is especially useful for encoding the input text which the model generation should be conditioned on. Next, the models are trained in a supervised manner using Sequence-to-Sequence (Seq2Seq) autoregressive modeling on SKILL input-output samples. Seq2Seq training is where the input of an input-output pair is given and the model is tasked with generating the output one token at a time. This uni-directional way of modeling is most useful for the generation procedure since the model only has access to previous tokens it has generated. Figure \ref{fig:training_objectives} shows examples of both modeling techniques.

Validation data were used to stop training once validation losses stop improving to avoid overfitting. We have added a sentence to make it more clear. During training, the weights checkpoint with the best validation loss at the end of an epoch was saved. This saved checkpoint is the one used for evaluation on the validation set using the BLEU metric. The training was stopped early after five epochs of no improvement in the validation loss. Models were trained for a maximum of 50 epochs. The best training strategy for each model type in terms of BLEU score \cite{bleu_base} was kept for the final evaluation of the test data split. Additional models trained using training strategies that gave surprising results were also selected for final evaluation (see Section \ref{results_training}).

\subsection{Evaluation of Synthesized SKILL Code}\label{sect:evaluation}
One of the most difficult aspects of code generation is evaluation. Other works on code generation\cite{chen2021evaluating} \cite{thakur2022benchmarking} manually created their programming problems with corresponding verification programs to evaluate their models. While this has the advantage of being able to measure the functional correctness of generated code, it also has two main disadvantages. Firstly, manually creating a functional correctness evaluation dataset is time-consuming. Secondly, manually creating data could introduce bias in the sense that only \textit{toy examples} of problems are written in the evaluation dataset which are not representative of real-world programming tasks. It is very difficult to create a functional correctness dataset from real-world repositories since (i) for a program to run as expected, the dependencies of the original file must be available and (ii) a corresponding verification problem must be available and linked to the code snippet. Instead, we rely on a static-analysis score and the popular BLEU metric \cite{bleu_base} to evaluate generated code.

\subsubsection{Static-Analysis Metric}

The static analysis used in this study was the change ($\Delta$) in the SKILL lint IQ score. As discussed in Section \ref{filtering}, SKILL lint IQ takes into account the syntactic correctness, style, and efficiency of SKILL programs. For a given input-output pair $p$ and model $m$, the change in lint IQ score is equal to $liq(f\vert m(p)) - liq(f)$ with $liq(f)$ the lint IQ score of the file $f$ which $p$ belongs to and $f\vert m(p)$ the file with the output of $p$ replaced by the model's prediction given the input of $p$.

\subsubsection{BLEU Implementation}
The implementation of the BLEU metric used in this study is the geometric mean of the $n$-gram precisions between the predicted sequence and reference sequences with $n\in\{1,2,3,4\}$. No smoothing was applied. This gives a score between 0 (bad) and 1 (good). Unless mentioned otherwise, the BLEU score was measured using the tokenizer of the respective model being evaluated. That is, the code generated from a model and the output reference code is tokenized using the tokenizer of the generation model and these two sets of tokens are compared. Post \cite{sacrebleu} argued that using different tokenizers has significant effects on BLEU scores and therefore BLEU scores calculated using different tokenizers were not compared. A standard tokenizer was chosen from the tokenizers of the different model types by measuring the correlation between their BLEU scores and human judgment scores to be able to use BLEU to compare the different model types.

\subsubsection{Correlation of Metrics with Human Judgement}
A small survey was developed to calculate the correlation between human judgment scores with the change in SKILL lint IQ metric and BLEU calculated with a SKILL-specific tokenizer. For a given input-output pair and output code predictions of all models being evaluated, the survey asked the following questions:
\begin{enumerate}
    \item ``On a scale of 1 (very bad) to 5 (very good), how would you rate the quality of the input prompt (i.e. how well did the input prompt describe the code it was prompting)?'' 
    \item ``On a scale of 1 (not at all) to 5 (very much), to what degree does the output reference follow logically from the input prompts?''
    \item ``On a scale of 1 (very bad) to 10 (very good), rate each model's output based on the quality of the generated SKILL code as well as the degree to which the generated SKILL code logically follows from the input prompt.''
\end{enumerate}
Additionally, the survey included open-ended questions for the survey taker to provide any additional feedback they wanted to provide. The survey was sent to SKILL developers who have access rights to view the proprietary data sources of our custom SKILL dataset (see Section  \ref{sect:prop_data}).

The survey included 15 SKILL code autocompletion input-output pairs (five of each pair type). The number of input-output pairs was limited to 15 to minimize the time required from human evaluators. The prompts were selected as follows: 
\begin{enumerate}
    \item Randomly sample 60 pairs of the corresponding pair type from the test dataset.
    \item Manually select the best 15 samples in terms of code quality and degree to which the output follows logically from the input.
    \item Randomly sample 5 of these 15 pairs for inclusion in the survey. The same prompts were used for all different models evaluated in the survey.
\end{enumerate}
This selection procedure was intended to achieve a balance between random sampling and ensuring that nonsensical prompts were not included in the survey. The latter is important due to the small number of pairs and model outputs that could be evaluated.

\section{Experimental Setup \& Pre-processing}\label{exp_setup}
All experiments were conducted on a workstation with an NVIDIA GeForce RTX 3070 graphical processing unit (GPU). The HuggingFace library \cite{huggingface} was used for many parts of the experiments including the implementations of the T5 \cite{t5}-based models\footnote{The pre-trained model checkpoints used for each model type were: \begin{itemize}
    \item CodeTrans: ``SEBIS/code trans t5 small api generation multitask''
    \item CodeT5: ``Salesforce/codet5-small''
    \item T5-NL: ``t5-small''
\end{itemize}}, BLEU \cite{bleu_base} metric, and parts of the training and inference scripts. The optimizer used was Adafactor \cite{adafactor}, the same optimizer used for the original T5 \cite{t5} models. The maximum batch sizes that fit into memory, 11 for self-supervised training and 2 for supervised training, were used.

For file data, duplicate files were removed (the number of files reported in Section \ref{results_discussion} is after removing duplicate files). Files were considered to be duplicates if their text values were the same. On the remaining files, unwanted metadata was removed. Unwanted metadata included liability disclaimers, author information (names and emails), and script version information. These types of metadata were found and removed using regular expressions. Characters from foreign characters, as well as other non-ASCII characters, were removed from all the files. Unnecessary comments were observed in the SKILL file dataset. Cases that were removed were commented-out code (based on the inclusion of keywords often found in commented-out code such as \textit{printf(}) and comments that contain only whitespace or only contain special characters (used predominately to separate files into segments). 

For input-output data samples, different pre-processing steps were applied to the input comment and output code part of the data, respectively. All single-line comments (for which the comment identifier is a semicolon and the comment extends until a newline character is encountered) were converted to multi-line comments (The body of the comment is surrounded by \textit{/*} and \textit{*/}). All comments in the output code were removed.

For MLM, the data was pre-processed by tokenizing, concatenating tokens from all SKILL files (separated by the \textit{end of sequence} special token), and creating sequences of 512 tokens from the full sequence of tokens. These sub-sequences were the training samples given to the models. For each training sample, a masking percentage of 15\% with a mean mask sequence length of 3 tokens was used, similar to how T5 \cite{t5} pre-processed unlabeled data. 

For Seq2Seq training and evaluation, a maximum input sequence length of 1024 and a maximum output sequence length of 512 were used. The input sequence length is longer because truncation of the input sequence is worse than truncation of the output sequence since the outputs are evaluated and are conditioned on the input sequences. This is especially the case if function definitions are truncated. To reduce the number of times input sequences have to be truncated, only the first 150 words (separated by whitespace) in comments of inputs are kept in the input sequence before truncation. This is similar to the strategy used for the CodeSearchNet dataset \cite{codesearchnet} where only the first paragraph in a documentation string is kept. Beam search with a beam size of 10 was used for all models during all evaluation procedures.

\section{Results \& Discussion}\label{results_discussion}
This section presents and discusses the experimental results of the methodology presented in Section \ref{methodology}. First, statistics of the custom SKILL dataset are shown in Section \ref{results_dataset}. Second, the validation results from training the pre-trained (or not in the case of T5-SKILL) models using different training strategies are shown in Section \ref{results_training}. This includes selecting the models for final evaluation by BLEU score. Finally, Section \ref{results_eval} presents the test results in terms of BLEU (using a SKILL-specific tokenizer), SKILL lint IQ, and human evaluation score.

\subsection{SKILL Dataset Statistics}\label{results_dataset}
\begin{figure}[ht]
    \centering
    \includegraphics[width=0.5\textwidth]{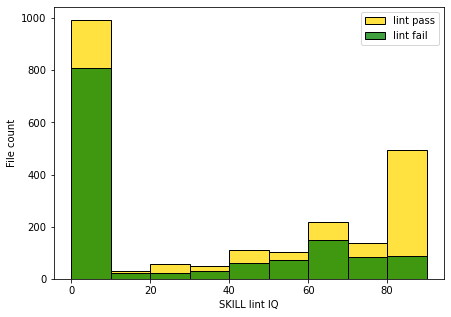}
    \caption{Stacked histogram of the SKILL lint IQ score of the training data files. Each bar is split by the number of files that received a ``pass'' or ``fail'' grade from the SKILL lint tool.}
    \label{fig:lint_histogram}
\end{figure}

The custom SKILL dataset consists of file samples for self-supervised training and input-output pairs for supervised training and evaluation. The numbers of proprietary, open-source, and total SKILL files in the dataset are 983, 2265, and 3248, respectively. Table \ref{tab:self-supervised_data} shows the number of different SKILL files by source, split, and filtering strategy. The lint pass criterion is the most strict, filtering out more than half of all training files. Figure \ref{fig:lint_histogram} shows a histogram for the SKILL lint results for each file. Filtering files using a minimum lint IQ score also removes many files. A  minimum lint IQ score of 10 was chosen because a large number of files had a lower score (see Figure \ref{fig:lint_histogram}). Manually sampling and inspecting files with a lint score smaller than 10 confirmed that many are either very short or not high-quality SKILL files. Requiring the presence of input-output pairs leaves the largest number of remaining files. The non-filtered dataset was mined to create the supervised dataset as described in Section \ref{sect:supervised_dataset}. Table \ref{tab:dataset_split} shows the number of input-output pairs by the dataset split. 

\begin{figure}[!t]
\centering
\subfloat[]{\includegraphics[trim={0 0 0 0},clip,width=0.48\textwidth]{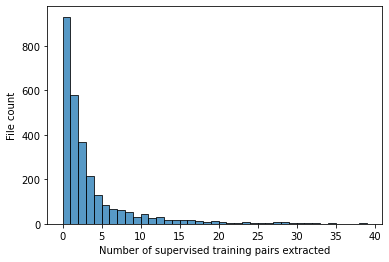}
\label{fig:0-40_hist}}
\hfil
\subfloat[]{\includegraphics[trim={0 0 0 0},clip,width=0.48\textwidth]{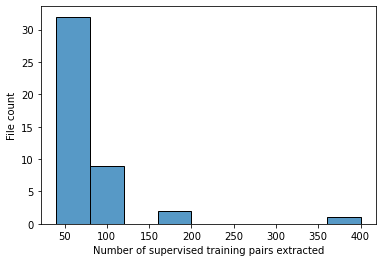}
\label{fig:40-400_hist}}
\caption{Histograms of training files by the number of supervised pairs extracted from them for 0-40 pairs (\ref{fig:0-40_hist}) and 40-400 pairs (\ref{fig:40-400_hist}). The histogram was split in two for visualization purposes.}
\label{fig:supervised_pairs_hist}
\end{figure}

Figure \ref{fig:supervised_pairs_hist} shows a histogram of the number of input-output pairs mined for each file. Four hundred pairs were mined from one outlier file. Only 18 of these 400 pairs were top-level pairs meaning the other 382 pairs were removed after pair deduplication. This could be a reason why deduplication was found to be effective for training as shown in Subsection \ref{results_training}.

\begin{table}[ht]
        \caption{Number of files in the custom SKILL dataset by source and split. The number of training files that remain after three different filtering methods is also shown.}
        \label{tab:self-supervised_data}
        \centering
        \begin{tabular}{|c|c||c|c|c|c|c|}
        \hline
        \multirow{2}{*}{\textbf{Split}}
         & \multirow{2}{*}{\textbf{Filtering}} & \multicolumn{2}{c|}{\textbf{Proprietary}} & \multicolumn{2}{c|}{\textbf{Open-source}} & \multirow{2}{*}{\textbf{All}} \\
        \cline{3-6}
         & & \textbf{Primary} & \textbf{Secondary} & \textbf{Repository search} & \textbf{Code search} &\\\hline\hline
            \multirow{4}{*}{Train} & - &  21 & 548 & 820 & 1445 & 2834 \\ \cline{2-7}
             & $\geq$ 1 pairs mined & 0 & 319 & 699 & 885 & 1903 \\ \cline{2-7}
             & lint IQ $\geq$ 10 & 20 & 223 & 654 & 934 & 1831 \\ \cline{2-7}
             & lint pass & 18 & 305 & 532 & 588 & 1443 \\ \hline
            Val & - & 207 & 0 & 0 & 0 & 207 \\ \hline
            Test & - & 207 & 0 & 0 & 0 & 207 \\ \hline
            All & - & 435 & 548 & 820 & 1445 & 3248 \\ \hline
        \end{tabular}
        
    \end{table}

\begin{table}[ht]
    \caption{Number of comment-function (CF), comment-code (CC), and function-completion (FC) pairs in the custom SKILL dataset by the split.}
    \label{tab:dataset_split}
    \centering
    \begin{tabular}{|c||c|c|c|c|}
        \hline
        \textbf{Split} & \textbf{CF} & \textbf{CC} & \textbf{FC} & \textbf{All} \\
        \hline \hline
        Train (All) & 1,212 & 6,766 & 4,564 & 12,542 \\
        \hline
        Train (Deduplicated)  & 1,200 &  2,303 & 2,495 & 5,998 \\ \hline
        Validation  & 180 & 180 & 180 & 540 \\
        \hline
        Test  & 177 & 177 & 177 & 531 \\
        \hline
    \end{tabular}
\end{table}

Table \ref{tab:dataset_comparison} compares the size of the custom SKILL dataset to the reported sizes of different subsets of the  CodeTrans \cite{codetrans} training dataset. The number of files in the smallest file subset is almost 53 times larger than the number of files in the SKILL dataset. The number of snippet-level samples in the smallest subset by total samples is almost 10 times larger than the number of input-output pairs in the SKILL dataset. This shows that the custom SKILL dataset is extremely small compared to other code datasets. This is the main motivation for the use of weights pre-trained on general PL data and fine-tuning on both unlabeled and labeled data to use the small SKILL dataset as efficiently as possible. Still, transformers \cite{attention_is_all} tend to overfit small datasets. Increasing the size of the SKILL dataset will be needed in future work as discussed in Section \ref{limitations}.

\begin{table}[ht]
    \caption{Size of the obtained SKILL training dataset compared with subsets of the dataset used to train CodeTrans. The subsets chosen are the two largest (Java and Python) and the four smallest (CSharp, Ruby, SQL, LISP) by the total number of samples categorized by PL.}
    \label{tab:dataset_comparison}
    \begin{tabular}{|c||>{\centering\arraybackslash}m{80pt}|>{\centering\arraybackslash}m{80pt}|
    >{\centering\arraybackslash}m{220pt}
    |}
    \hline
        \textbf{Language} & \textbf{File-level samples} & \textbf{Snippet-level samples} & \textbf{Sources} \\ \hline \hline
        Java & 720,124 & 9,581,115 & CodeSearchNet \cite{codesearchnet}, CodeNN \cite{codenn}, DeepCom \cite{deepcom}, DeepAPI \cite{deepapi}, Public Git Archive \cite{public_git_archive} \\ \hline
        Python & 149,114 & 1,296,064 & CodeSearchNet \cite{codesearchnet}, CodeNN \cite{codenn}, Python150K \cite{python150}
         \\ \hline
        CSharp & 469,038 & 52,943 & CodeNN \cite{codenn}, Public Git Archive \cite{public_git_archive} \\ \hline
        Ruby & 0 &  179,281 & CodeSearchNet \cite{codesearchnet} \\ \hline
        SQL & 0 &  158,862 & CodeNN \cite{codenn}, StaQC \cite{staqc} \\ \hline
        LISP & 0 & 122,602 & GitHub \cite{github_website}\\ \hline \hline
        SKILL & 2,834 & 12,542 & Proprietary sources, GitHub \cite{github_website}\\ \hline
    \end{tabular}
\end{table} 

\subsection{Validation Results}\label{results_training}
The proposed models and the two baselines were trained and validated on the custom SKILL dataset. For each model type (CodeTrans \cite{codetrans}, CodeT5 \cite{codet5}, T5-NL \cite{t5}, and T5-SKILL), all training strategies that can be made through all possible combinations of (un-)supervised learning and the filtering and deduplication strategies described in Section \ref{filtering} were applied to train a model. 

Table \ref{tab:val_results} shows the results of the top three models in terms of BLEU score. Interestingly, both baselines achieved better BLEU scores without supervised training. It is believed that these results come from degenerated n-gram repeating and repeating tokens from the input which result in deceptively high BLEU scores. The best training strategy for each of these models which were trained in a supervised manner are also shown in Table \ref{tab:val_results} and these models were selected for final evaluation on the test set for further investigation. Another training strategy that was not the best of its model type but was chosen for evaluation on the test set was the CodeTrans model which was not fine-tuned in a self-supervised manner. This model was the third best CodeTrans model trained based on validation BLEU score. This was surprising since all other models pre-trained on general PL data do seem to benefit from self-supervised fine-tuning, based on the BLEU validation results. To investigate further, this model was selected for final evaluation on the test set. 

\begin{table}[ht]
    \caption{Evaluation results for the top three models, and interesting outliers, in terms of BLEU score on the validation dataset split. Here we denote a model as a model type trained using a certain training strategy (columns 3-7). The rank of a model in terms of BLEU score (final column) relative to other models of the same model type is shown in the second column. Models that were selected for final evaluation on the test set have * next to their rank.} 
    \label{tab:val_results}
    \centering
    \begin{tabular}{|>{\raggedright\arraybackslash}c||c|c|c|c|c|c|c|}
        \hline
        \multirow{2}{*}[-1.5em]{\textbf{Model type}} & 
        \multirow{2}{*}[-1.5em]{\textbf{Rank}} & 
        \multicolumn{3}{|c|}{Self-supervised training} & 
        \multicolumn{2}{|c|}{\textbf{Supervised}} & \multirow{2}{*}[-1.5em]{\textbf{BLEU}} \\
        \cline{3-7}
         & & \rotatebox[origin=c]{90}{\hspace{1pt} \textbf{applied?} \hspace{1pt}} & \textbf{file filtering?} & \textbf{comments?} & \rotatebox[origin=c]{90}{\hspace{1pt} \textbf{applied?} \hspace{1pt}} & \textbf{deduplicated?} & \\
        \hline \hline
        \multirow{3}{*}{CodeTrans} & 1* & \cmark & lint pass & \cmark & \cmark & \cmark & 0.0864 \\
        \cline{2-8}
         & 2 & \cmark & $\geq$ 1 pairs mined & \cmark & \cmark & \cmark & 0.0739 \\ \cline{2-8}
         & 3* & \xmark & NA & NA & \cmark & \cmark & 0.0676 \\ \hline
        \multirow{3}{*}{CodeT5} & 1* & \cmark & lint IQ $\geq$ 10 & \xmark & \cmark & \cmark & 0.0470 \\ \cline{2-8}
         & 2 & \cmark & - & \xmark & \cmark & \cmark & 0.0465 \\ \cline{2-8}
         & 3 & \cmark & - & \cmark & \cmark & \cmark & 0.0455 \\ \hline
        \multirow{4}{*}{T5-NL} & 1* & \cmark & - & \xmark & \xmark & NA & 0.0623 \\ \cline{2-8}
         & 2 & \cmark & lint pass & \cmark & \xmark & NA  & 0.0619 \\ \cline{2-8}
         & 3 & \cmark & lint pass & \xmark & \xmark & NA  & 0.0567 \\ \cline{2-8}
         & 4* & \cmark & $\geq$ 1 pairs mined & \cmark & \cmark & \cmark & 0.0522 \\ \hline
        \multirow{4}{*}{T5-SKILL} & 1* & \cmark & - & \xmark & \xmark & NA & 0.0829 \\ \cline{2-8}
         & 2 & \cmark & lint IQ $\geq$ 10 & \xmark & \xmark & NA & 0.0820 \\ \cline{2-8}
         & 3 & \cmark & lint IQ $\geq$ 10 & \cmark & \xmark & NA & 0.0737 \\ \cline{2-8}
         & 9* & \cmark & $\geq$ 1 pairs mined & \cmark & \cmark & \cmark & 0.0376 \\ \hline
    \end{tabular}
\end{table}

\begin{figure}[ht]
    \includegraphics[width=\textwidth]{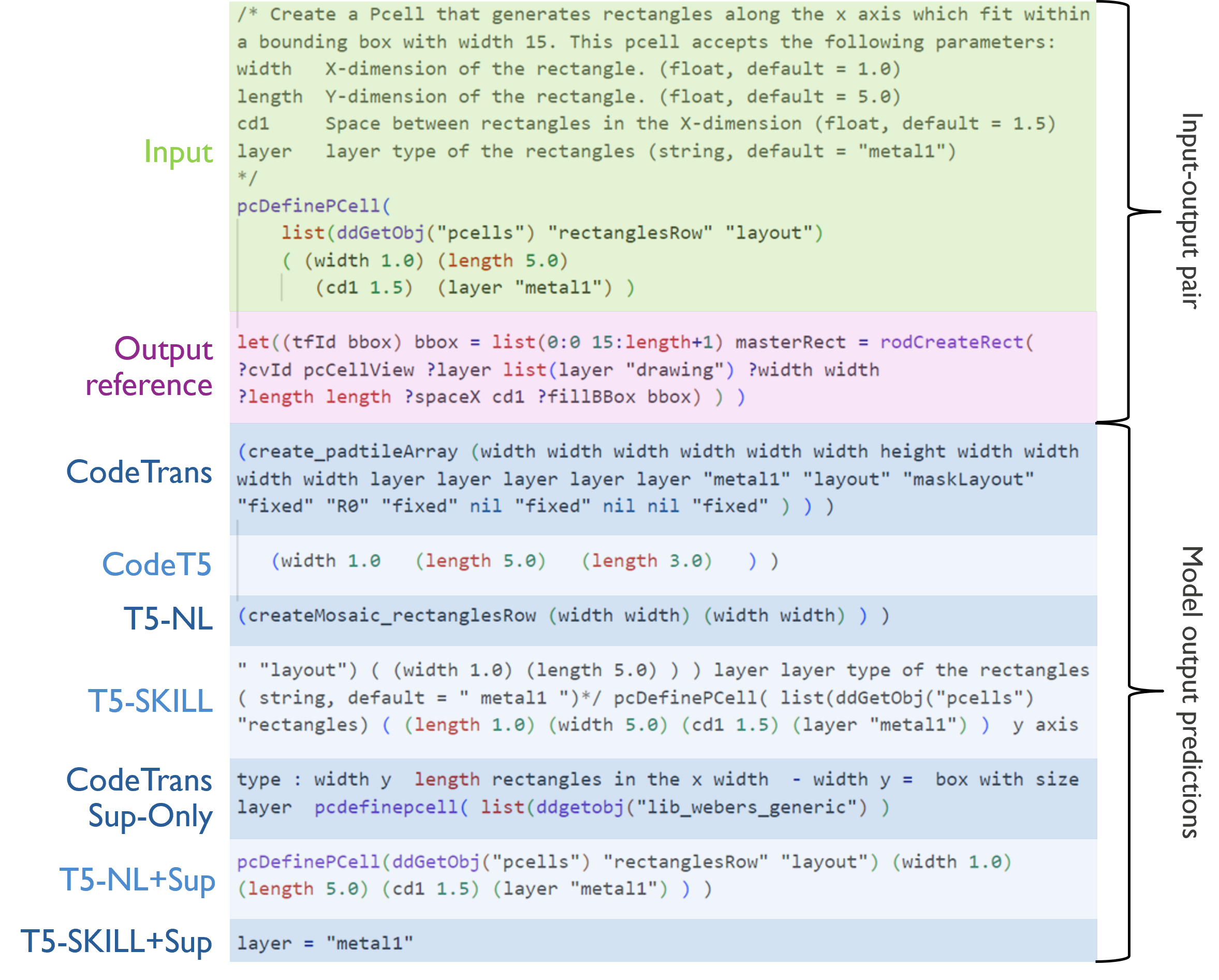}
    \caption{Model predictions for the example input prompt of the program on the left of Figure \ref{fig:example_input_output}. Note that the output reference has been formatted to save space. Newlines were manually added to the model outputs to save space and improve readability.}
    \label{fig:model_example_outputs}
\end{figure}

Figure \ref{fig:model_example_outputs} shows an example output for each model selected for final evaluation. Ideally, these models were to output SKILL code that is functionally equivalent to the output reference. These outputs demonstrate that the models are not yet capable enough to output compilable SKILL programs. Reasons for this and how SKILL code generation performance can be improved are discussed in Section \ref{limitations}. The output of the best CodeTrans model is an example of repeating tokens, namely the ``width'' and ``layer'' tokens. The output of the best T5-NL model is an example of repeating tokens from the input. The differences in the model outputs despite all the models being given the same input show that the pre-training data, fine-tuning strategy, and tokenizers affect their generation capabilities.

\begin{figure}[!t]
\centering
\subfloat[]{\includegraphics[width=0.93\textwidth]{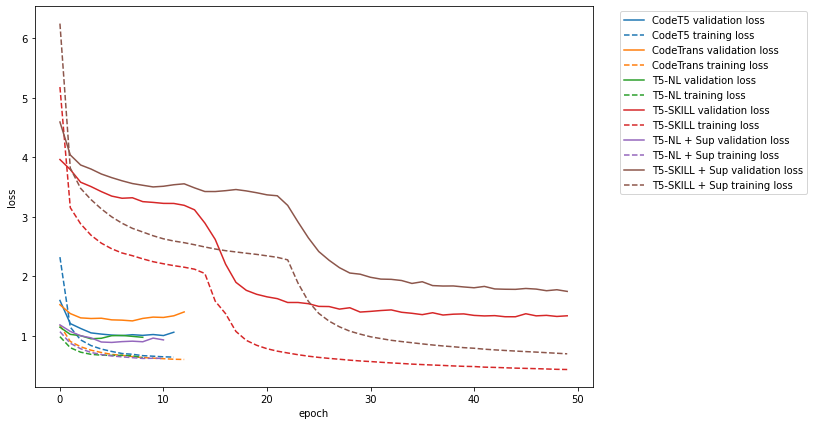}
\label{fig:unsup_losses}}
\hfil
\subfloat[]{\includegraphics[width=0.8\textwidth]{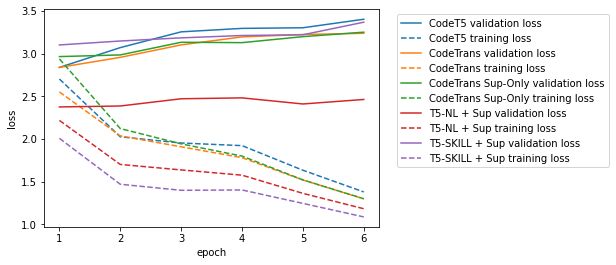}
\label{fig:sup_losses}}
\caption{Self-supervised (\ref{fig:unsup_losses}) and supervised (\ref{fig:sup_losses}) training and validation loss curves for models selected
for final evaluation.}
\label{fig:losses}
\end{figure}

All the best models that were trained in a supervised manner were trained on the deduplicated supervised training set. The top models trained in a supervised manner for every model type used file-filtering for self-supervised training with the results suggesting that there was a correlation between the size of the pre-training code dataset and the amount of unlabeled data that is filtered. For example, the top CodeTrans \cite{codetrans} model used the largest code dataset for pre-training and uses the filtering technique that filters the most number of SKILL files, the SKILL lint pass grade. Whether including comments in the self-supervised training set helps the models learn is inconclusive from the results obtained.

Figure \ref{fig:losses} shows the training and validation loss curves for the models selected for final evaluation. The validation loss curves during self-supervised learning (Figure \ref{fig:unsup_losses}) of pre-trained models start to saturate within the first 7 epochs of training. The models trained from scratch do not saturate even after 50 epochs and have higher validation losses than the pre-trained models. This suggests that pre-training helps the model learn the SKILL language faster and better. The validation loss curves for supervised learning (Figure \ref{fig:sup_losses}) do not improve after the first epoch. This suggests that the supervised learning task is not as effective for learning to model the SKILL language as it quickly starts to overfit. However, as the next section will show, supervised training is still beneficial in terms of BLEU score and human judgment score for models pre-trained on general PL data.

\subsection{Final Evaluation Results}\label{results_eval}
\begin{figure}[!t]
\centering
\includegraphics[width=0.5\textwidth]{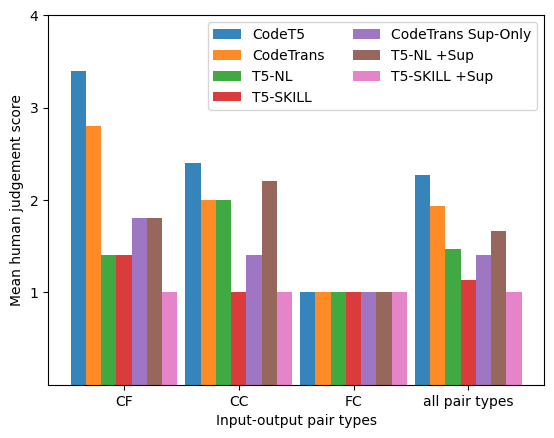}
\caption{Mean human evaluation scores for the different types of pairs as well as the overall mean score for all pair types. CF, CC, and FC stand for comment-function, comment-code, and function-completion pairs, respectively.}
\label{fig:human_eval_bars}
\end{figure}

The selected models from Table \ref{tab:val_results} were evaluated on the test split of input-output pairs. Figure \ref{fig:human_eval_bars} shows the scores assigned to each model for each sample in the human evaluation study and the mean scores for each pair type. Function-completion pairs were given a score of 1 for every sample. The survey results gave mean scores of 4.4 and 4.2 out of 5 for the quality of the input prompt and output references, respectively, for the function-completion samples so the comprehensiveness of the function-completion pairs is most likely not the problem. The results for comment-function and comment-code pairs were better for most of the models. Both T5-SKILL models performed poorly which supports the hypothesis that the custom SKILL dataset is too small to train a model from scratch. These results also suggest that the reason for the poor results on the function-completion pairs might be because the pre-trained models have been trained on a lot of NL and so this knowledge has transferred much better than the models have been able to learn the SKILL language. 

The CodeT5 model \cite{codet5} achieved the highest overall mean score of 2.27. The CodeTrans model \cite{codetrans} achieved the next-best mean score of 1.93. The T5-NL model \cite{t5} with supervised training achieved a mean score of 1.67, a better score than the T5-NL model not trained in a supervised manner and the CodeTrans model without self-supervised training. This suggests that both self-supervised and supervised training are beneficial. When asked to score whether the best outputs for each sample would be useful if the evaluator was tasked with writing SKILL code for the corresponding input prompts, the human evaluator gave scores of 1, 2, and 1 out of 5 for comment-function, comment-code, and function-completion pairs, respectively. This refutes the idea that the models trained on the custom SKILL dataset are ready to be deployed in a real-world setting to assist SKILL developers. 

\begin{figure}[ht]
    \centering
    \includegraphics[width=\textwidth]{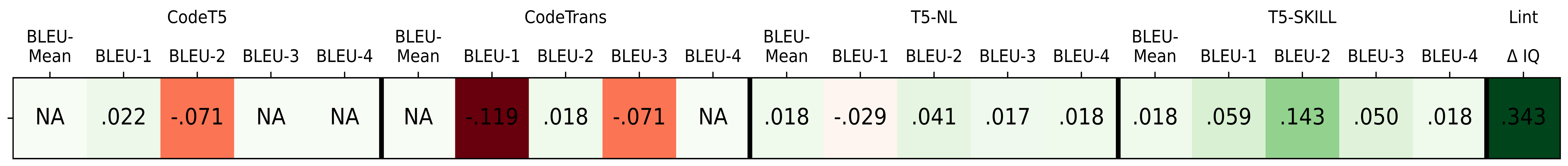}
    \caption{Pearson correlation coefficients of BLEU variants using different tokenizers (corresponding to the four model types) and the change in lint IQ score with comment-function and comment-code human evaluation scores. The BLEU-n scores are the BLEU scores when only considering n-gram precision. ``BLEU-Mean'' denotes the same BLEU score used in the rest of the paper (geometric mean of the 1,2,3,4-gram precisions).}
    \label{fig:correlation}
\end{figure}

The two automatic evaluation metrics, BLEU using different tokenizers and lint IQ, were measured on the same samples from the human evaluation survey to calculate the correlation between them. Figure \ref{fig:correlation} shows that the correlation between the change in the lint IQ score correlates well with human scores compared to BLEU scores. Since the function-completion human-assigned scores were all the same, these results are counter-productive for the calculation of the correlations and have not been included in the calculations of the coefficients. The high correlation of the changes in SKILL lint IQ with human evaluation scores might be attributable to the overall low quality of code synthesized by the models. Synthesized code that was syntactically and semantically correct was probably judged relatively favorably by human evaluators since little to none of the generated code followed logically from the input prompt. That is, the functional correctness of the SKILL code generated from all the models remained consistently low.

For BLEU scores, the best correlations are achieved when the monotonic function-completion scores of the human evaluation are ignored. As the models improve, especially by reducing the frequency of degenerate token copying and repetition (feedback received from the open-ended questions on the human judgment survey), BLEU scores should correlate better with human evaluation scores. The tokenizer that achieved the best correlation between BLEU and human scores is the T5-SKILL tokenizer. This was to be expected since this tokenizer was trained on the custom SKILL dataset and therefore has a vocabulary that is most relevant for SKILL. Therefore, the T5-SKILL tokenizer was used for the standardized BLEU score. Surprisingly, the T5-NL tokenizer \cite{t5} correlated better with human scores than the tokenizers of the proposed models which suggests that general PL tokenizers might not be very relevant for the SKILL PL.

The final results on the entire test set (except for human evaluation scores) for the selected models are shown in Table \ref{tab:test_results}. The CodeTrans model \cite{codetrans} that achieved the best validation BLEU score also achieved the best SKILL standardized BLEU score of 0.032. This is double or more than what all non-CodeTrans models achieve. CodeTrans (with and without self-supervised training), CodeT5 \cite{codet5}, and T5-NL \cite{t5} with supervised training achieved similar mean changes in lint IQ scores. As discussed above, CodeT5 performed best in the human evaluation with CodeTrans performing second best. These results support the hypothesis that models pre-trained on large volumes of PL data can more efficiently learn unseen PLs than models pre-trained only on NL or models trained from random initial weights. The results also support the hypothesis that fine-tuning using both supervised and self-supervised learning improves SKILL code generation models.

However, the overall results of the models indicate that the models are not yet capable of reliable SKILL code generation. If the models were generating quality SKILL code, we could expect the mean change in lint IQ to be in the range of $-30$ and $-10$, indicating mostly style and efficiency errors instead of significant syntax and compilability errors. Furthermore, the mean human scores should be at least 5 to indicate that most generated code snippets are at least plausible to be correct SKILL statements to a SKILL developer. The next section discusses the most likely reasons for these poor results. The BLEU score results are harder to interpret without relating to baselines. Subsection \ref{lim_sub_eval} discusses BLEU's viability as a SKILL code generation metric for future works.

\begin{table}[ht]
    \caption{Final evaluation results on the input-output pair test split of the custom SKILL dataset for different models in terms of BLEU, T5-SKILL-BLEU (BLEU score calculated using the T5-SKILL tokenizer), and change in SKILL lint IQ metric. The mean human evaluation scores for each model for 15 pairs (see Section \ref{sect:evaluation}) are provided in the final column.}
    \label{tab:test_results}
    \centering
    \begin{tabular}{|c||c|c|c|c|}
        \hline
        \textbf{Model} & \textbf{BLEU} & \textbf{T5-SKILL-BLEU} & \textbf{Mean $\Delta$ Lint IQ} & \textbf{Mean Human Score}\\ \hline \hline
        CodeTrans & \textbf{0.019} & \textbf{0.032} & \textbf{-60.7} & 1.93 \\ \hline
        CodeT5 & 0.008 & 0.015 & \textbf{-60.3} & \textbf{2.27} \\ \hline
        T5-NL & 0.004 & 0.007 & -70.0 & 1.46 \\ \hline
        T5-SKILL & 0.016 & 0.016 &  -71.7 & 1.13 \\ \hline \hline
        CodeTrans Sup-Only & 0.014 & 0.024 & \textbf{-59.4} & 1.40 \\ \hline
        T5-NL +Sup & 0.002 & 0.002 & \textbf{-59.8} & 1.67 \\ \hline
        T5-SKILL +Sup & 0.002 & 0.002 & -64.4 & 1.00 \\ \hline 
    \end{tabular}
    
\end{table}

\section{Limitations \& Proposed Future Work} \label{limitations}
The results of the experiments discussed in the previous section suggest that more work must be done before the proposed methodology can be deployed to assist SKILL developers. In this section, the main limitations of the proposed methodology are discussed. These limitations are dataset scaling and confidentiality, size of the models, and evaluating SKILL code. For each limitation, promising research directions for future work that address these limitations are proposed.

\subsection{Dataset Scaling \& Confidentiality}\label{lim_sub_scaling}
Table \ref{tab:dataset_comparison} showed that the size of the custom SKILL dataset is orders of magnitude smaller than datasets of other PLs. The amount of open-source SKILL data is small relative to general PLs such as Python \cite{python_software} or Java \cite{java_software} largely because the SKILL language itself is proprietary. While there are a large number of proprietary SKILL repositories, a very limited number of them were able to be used for the experiments in this study due to confidentiality concerns. Future work in privacy-preserving ML, such as homomorphic encryption \cite{homomorphic_survey} or federated learning \cite{federated_learning_survey}, could alleviate these confidentiality concerns and allow for data collection across teams and organizations. 

GitHub data was not checked for permissive licenses. For this study, this should not be a problem since the code was only forked and not further distributed nor was it profited from. Furthermore, O'Keefe et al. \cite{openai_copyright_github} argued that training ML models on copyrighted works should constitute fair use but there is still legal ambiguity. Even if there are no legal issues, the use of free and open-source software to create proprietary products presents an ethical concern for the developers of such software \cite{foss_leaves_github}.

\subsection{Model Size}\label{lim_sub_model_size}
All the models trained in this study are based on the ``small'' T5 architecture \cite{t5}. Chowdhery et al. \cite{palm} showed that increasing the size of a language model not only generally improves performance but that certain tasks require models with a certain minimum number of parameters to obtain acceptable results. The relationship between the inputs and outputs of the supervised dataset is complex and a higher number of parameters might be necessary to model it effectively. Small architectures were used due to a combination of limited computation resources and a large number of different models trained (72).

Future work could use the results from this study to limit the model search space allowing for fewer but larger models to be explored. However, larger models require more computing resources which can lead to higher energy consumption and in turn leads to high financial costs and negative climate change effects  \cite{how_big_is_too_big}. Additionally, high computing requirements usually lead to the centralization of the model on a server where client nodes can send requests too. This increases the risk of breaches in data privacy which is important to the semiconductor industry as discussed in the previous subsection.

\subsection{Evaluating SKILL Code}\label{lim_sub_eval}
Despite its popularity, Ren et al. \cite{codebleu} argue against using metrics intended for the evaluation of NL (such as BLEU) for evaluating code. This is because functionally equivalent programs can consist of different tokens and BLEU does not take this into account. Instead, Ren et al. \cite{codebleu} proposes CodeBLEU which is inspired by the original BLEU metric \cite{bleu_base} which is more appropriate for evaluating synthesized code. CodeBLEU uses a weighted sum of BLEU, BLEU weighted on code keywords, syntactic similarity by comparing ASTs, and data-flow similarity. CodeBLEU could not be used for the SKILL code generation experiments since the SKILL parser is not publicly available and so there was no simple way to obtain ASTs of SKILL code. Since the correlation between changes in SKILL lint IQ scores and human judgments was shown to be relatively strong in Section \ref{results_eval}, future work should investigate how well a combination of BLEU and SKILL lint IQ correlate with human judgments. 

SKILL lint IQ is an automatic metric that requires no reference code. A promising direction for future work would be to sample multiple outputs from the models and use the lint IQ score to choose the best output as the final prediction. Multiple outputs can be sampled from language models through generation strategies such as top-k sampling \cite{top_k_sampling}. This is inspired by, but different from, related works that achieve improved functional correctness metrics by sampling a large number of outputs and relying on unit tests to filter candidate predictions \cite{chen2021evaluating, alphacode}. 

When a model is developed that can reliably generate SKILL code with high SKILL lint IQs, the functional correctness of the generated code can be evaluated.  This requires verification programs for each test program. For physical layout generation, this will require PCells to be paired with a list of parameter values and corresponding physical layout instantiations. Preferably, programs to measure functional correctness should be mined from real-world data instead of manually writing SKILL code for functional correctness. This is discussed in more detail in Section \ref{sect:evaluation}.

The sample size of the human evaluation study was small, with 15 different prompts for each model. This means that a high variance of the human evaluation results can be expected were the study to be replicated using different prompts or a different human evaluator. Future work should ensure that human evaluations of SKILL code are easier or more accessible to allow for more samples to be evaluated.

\section{Conclusion}\label{conclusion}
Token-based code generation using deep learning transformer models promises to improve the productivity of programmers. These transformer models generally require large code datasets to learn to model a programming language effectively. In this work, we present the first study on transformer-based SKILL code autocompletion towards improving the productivity of design engineers facing increasingly more complex hardware design challenges. We propose a novel, data-efficient methodology for training models to autocomplete SKILL code. This methodology includes the creation of a high-quality SKILL dataset that contains unlabeled and labeled data. It also includes fine-tuning T5-based models pre-trained on general PL data on both the unlabeled and labeled data. 

Validation results suggested that applying a file quality filter to the self-supervised training set and deduplicating pairs in the supervised training set leads to improved BLEU scores. A human evaluation study was conducted and showed that BLEU scores obtained using a standard, SKILL-specific tokenizer correlated better to human judgments than using other tokenizers. It also showed that changes in SKILL lint IQ, a static analysis score based on syntactic and semantic correctness, correlated relatively well with human judgments. Test results showed that using models pre-trained on general PL data and fine-tuned using both unlabeled and labeled SKILL data improved BLEU, SKILL lint IQ, and human evaluation scores. However, the test results refute the viability of the trained models for reliable SKILL code autocompletion. Limitations of the study and corresponding suggestions for future work to address these limitations are discussed. Ultimately, the experimental results obtained provide valuable insights towards autocompleting SKILL code in a data-efficient manner. 

\section*{Acknowledgments}
We would like to thank our colleagues at imec, Victoria Malacara and Dr. Yasser Sherazi, for providing guidance and assistance with the SKILL programming language.

\bibliographystyle{spiebib}
\bibliography{references}

\end{document}